\documentclass[12pt]{article}
\pdfoutput=1 
\usepackage{color}
\definecolor{myblue}{rgb}{.8, .8, 1}

\usepackage{amsmath}   
\usepackage{empheq}
\usepackage{amsfonts}
\usepackage[bbgreekl]{mathbbol}
\makeatletter
\def\amsbb{\use@mathgroup \M@U \symAMSb}
\makeatother
\usepackage[bbgreekl]{mathbbol}
\usepackage{amssymb,amsfonts}

\usepackage{graphicx}
\usepackage{subfigure,psfig}

\newlength\mytemplen
\newsavebox\mytempbox

\makeatletter
\newcommand\mybluebox{%
    \@ifnextchar[
       {\@mybluebox}%
       {\@mybluebox[0pt]}}

\def\@mybluebox[#1]{%
    \@ifnextchar[
       {\@@mybluebox[#1]}%
       {\@@mybluebox[#1][0pt]}}

\def\@@mybluebox[#1][#2]#3{
    \sbox\mytempbox{#3}%
    \mytemplen\ht\mytempbox
    \advance\mytemplen #1\relax
    \ht\mytempbox\mytemplen
    \mytemplen\dp\mytempbox
    \advance\mytemplen #2\relax
    \dp\mytempbox\mytemplen
    \colorbox{myblue}{\hspace{1em}\usebox{\mytempbox}\hspace{1em}}}

\makeatother

\usepackage{bbm}
\usepackage{yfonts}
\pdfoutput=1
\usepackage{amsfonts}
\usepackage[bbgreekl]{mathbbol}
\makeatletter
\def\amsbb{\use@mathgroup \M@U \symAMSb}
\makeatother
\usepackage[bbgreekl]{mathbbol}
\usepackage{amssymb,amsfonts}
\usepackage{graphicx}
\usepackage[top=2.75cm, bottom=2.75cm, left=2.75cm, right=2.75cm]{geometry} 
\usepackage{appendix}
\usepackage{color}
\usepackage{caption}
\usepackage{lipsum}

\newcommand{\be}{\begin{equation}}
\newcommand{\ee}{\end{equation}}
\newcommand{\ber}{\begin{eqnarray}}
\newcommand{\eer}{\end{eqnarray}}

\def\+{{+\!\!\!+}}

\newcommand{\rd}[1]{{\color{red}{#1}}}

\newcommand{\nn}{\nonumber}
\newcommand{\pa}{\partial}
\newcommand{\na}{\nabla}
\newcommand{\half}{{\textstyle{\frac12}}}

\newcommand{\bbD}[1]{\mathbb{D}_{#1}}
\newcommand{\bbDB}[1]{\bar{\mathbb{D}}_{#1}}

\newcommand{\re}[1] {(\ref{#1})}

\numberwithin{equation}{section}

\begin{document}
\begin{titlepage}
\begin{flushright} \small
UUITP-14/16\\
Imperial-TP-2016-CH-01\\
Imperial-TP-2016-UL-01\\
\end{flushright}
\smallskip
\begin{center} 
\LARGE
{\bf All $(4,1)$: 
 Sigma Models with $(4,q)$ Off-Shell Supersymmetry} \\[30mm] 
\large
{\bf Chris Hull$^a$}~and~{\bf Ulf~Lindstr\"om$^{ab}$} \\[20mm]
{ \small\it
$^a$The Blackett Laboratory, Imperial College London\\
Prince Consort Road London SW7 @AZ, U.K.\\
$^b$ Department of  Physics and Astronomy,\\ Division of Theoretical Physics,
Uppsala University,\\ Box 516, SE-751 20 Uppsala, Sweden }
\end{center}

\vspace{10mm}
\centerline{\bfseries Abstract} 
\noindent
Off-shell $(4,q)$ supermultiplets in  2-dimensions are constructed for $q=1,2,4$. These are used to construct sigma models whose target spaces are hyperk\"ahler with torsion. The off-shell supersymmetry implies the three complex structures are simultaneously integrable and allows us to  construct  actions using extended   superspace and projective superspace, giving an explicit construction of the target space geometries.
\bigskip

\end{titlepage}

\tableofcontents
\section{Introduction}

There is a rich interplay between supersymmetry and geometry in non-linear sigma models. Supersymetric sigma models have led to the discovery of a rich class of complex geometries.
Our purpose here is to revisit this story, and we find that  past results readily extend to the construction of some interesting new geometries with $(4,q)$ supersymmetry.

The sigma model in 2 dimensions with $(1,1)$   supersymmetry has a target space with  a metric $g$ and closed 3-form $H$ given locally in terms of a 2-form potential $B$, $H=dB$  \cite{Gates:1984nk}.
The action can be written in $(1,1)$   superspace with coordinates 
$(x^{\mu},\theta^\alpha)$ 
where
$x^\mu = (x^\+,x^=)$ are null coordinates, $x^\+=\tau+\sigma$, $x^-=\tau-\sigma$,
  and $\theta ^\alpha =( \theta^+, \theta ^-)$.
 We use spinor indices $+,-$ so that $\psi ^+$ is a positive chirality 1-component Weyl spinor and $\psi ^-$ is a left-handed one, for any spinor $\psi$.
If the target space coordinates are $X^i$, $i=1,...,n$, the map from the worldsheet superspace to the target space is given locally by scalar superfields $X^i(x,\theta)$ and the action is %
\ber
S=\frac 1 2 \int d^2x 
d^2 \theta \,
D_-X^i(g+B)_{ij}(X)D_+X^j~.
\eer

For particular  geometries, the sigma model can have extended supersymmetry.
The conditions for $(2,2)$ and (4,4) supersymmetry were found in \cite{Gates:1984nk}
and the conditions for (2,0) supersymmetry were found in \cite{Hull:1985jv}.
This was generalised to the case of $(p,q)$ supersymmetry in \cite{Hull:1986hn} and
the geometry was further studied in \cite{Howe:1988cj}.
The $(1,1)$   theory will in fact have $(p,q)$ supersymmetry (with $p,q=1,2$ or $4$) if the target space has $p-1$ complex structures $J_{(+)}$ and $q-1$ complex structures $J_{(-)}$
satisfying
\ber
J^t_{(\pm)}gJ_{(\pm)}=g~,~~~(J_{(\pm)})^2=-\mathbb{1}~,~~~\na^{(\pm)}J_{(\pm)}=0~,
\eer
where 
\ber
\na^{(\pm)}:=(\na^{(0)}\pm\half g^{-1}H)
\eer
is the connection with torsion $\pm \frac 1 2 g^{il}H_{ljk}$ added to the  Levi-Civita connection
$\na^{(0)}$.
Then the extra supersymmetry transformations are given in terms of these   complex structures by
\ber\label{nis1tfs}
\delta X^i=\epsilon_A^+\left(J_{(+)}^{(A)}\right)^i_{~j}D_+X^j+\epsilon_{\tilde A}^-\left(J_{(-)}^{(\tilde A)}\right)^i_{~j}D_-X^j~,
\eer
where $A=1,...,p\!-\!1$ and $\tilde A=1,...,q\!-\!1$ label the  complex structures. 

 Closure of the algebra requires that    $J^{(A)}$ is an almost complex structure, $(J^{(A)})^2=-\mathbb{1}$ and that it is integrable, i.e. the Nijenhuis tensor vanishes, ${\cal N}(J^{(A)})=0$, so that it is a complex structure. Similarly, the $J^{(\tilde A)}$ are also complex structures.
When $p>1$ and/or $q>1$, the commutator   of supersymmetries $[\delta _{\epsilon_A}  ,\delta _{\epsilon_B} ]$ gives a term with 
involving a tensor  ${\cal N}(J^{(A)}, J^{(B)})$ constructed from the complex structures, known as
the Nijenhuis concomitant,
so that  for closure it is necessary that this vanishes.
For  three anticommuting almost complex structures $I,J,K$ satisfying the algebra of the quaternions it was   shown in  \cite{Yano:1972} that the vanishing of the Nijenhuis tensor of any two of the complex structures implies the vanishing of that of the third, and of all of the concomitants, so the integrability of the three complex structures $J^{(A)}$ is sufficient for closure, and in particular implies the vanishing of the Nijenhuis concomitant.

In what follows we shall be particularly interested in cases when there is a coordinate system (atlas) for which all the complex structures are constant in all coordinate patches, i.e. they are simultaneously integrable.
Three anticommuting complex structures  $I,J,K$  are simultaneously integrable when a certain curvature formed from the three of them vanishes $R(I,J,K)=0$ \cite{Obata 56, Yano:1973, Howe:1988cj}. 
For two complex structures, $J^{(+)}$ and $J^{(-)}$ that commute, $[J^{(+)},J^{(-)}]=0$, it is instead the vanishing of the Magri-Morosi concomitant, ${\cal M}(J^{(+)},J^{(-)})$ that signals simultaneous integrability. For details see \cite{Howe:1988cj}.

If $H=0$, then there are equal numbers of left and right handed supersymmetries,  $p=q$, and the target space is K\"ahler for $(2,2)$ supersymmetry and hyperk\"ahler for (4,4) supersymmetry.
For the $(2,2)$ case, the supersymmetry  algebra closes off-shell and the theory can be formulated in terms of chiral superfields, while for (4,4) supersymmetry, the supersymmetry algebra closes on-shell only, or after introducing an infinite number of auxiliary fields\footnote{E.g. using projective or harmonic superspace.}, as the 3 complex structures are not simultaneously integrable.

For $H\ne 0$, there is a richer structure. For $(2,1)$  supersymmetry, the supersymmetry  algebra closes off-shell and the theory can be formulated in terms of chiral superfields, while for $(2,2)$ supersymmetry the 
supersymmetry  algebra closes off-shell  only  once suitable auxiliary fields are introduced.
The theory can then be formulatd in terms of chiral superfields, twisted chiral superfields,  and semi-chiral superfields \cite{Lindstrom:2005zr}.
For  $(4,q)$   supersymmetry, the 
supersymmetry algebra only closes on-shell in general, but there are interesting cases in which the algebra closes off-shell, and the three complex structures $J^{(+)}$ are simultaneously integrable, i.e. there is a coordinate system where all of them are constant \cite{Howe:1988cj}.
One example of this is the (4,4) supersymmetric model found in  \cite{Gates:1984nk} that generalises that obtained  from the dimensional reduction of $N=2$ super-Yang-Mills theory in 4 dimensions.
The aim of this paper is to investigate such cases with off-shell $(4,q)$ supersymmetry, with simultaneously integrable complex structures $J^{(+)}$.
In such cases, there is an off-shell superfield formulation, and a superspace formulation of the action that gives a general local construction of the geometry in terms of  
certain potentials.
In this paper, a number of new multiplets will be found and analysed.
Actions for these multiplets will then be constructed using projective superspace. Projective superspace has a long history
 \cite{Karlhede:1984vr, Lindstrom:1987ks, Buscher:1987uw, Lindstrom:1994mw, Lindstrom:1989ne, GonzalezRey:1997qh, Lindstrom:2008gs} 
 paralleling and complementing  that of harmonic superspace \cite{Galperin:1984av}, \cite{Galperin:1985an}, 
 \cite{Delduc:1989gx} \cite{Galperin:2001uw}. General superspaces of this type have been described in \cite{Howe:1995md}, \cite{Hartwell:1994rp}. For detailed reviews of projective $(4,4)$ superspace see \cite{Lindstrom:2008gs} and the lectures \cite{Kuzenko:2010bd}.

The plan of the paper is as follows. In Section 2  we define an off-shell  $(4,1)$  multiplet that will play a key role in what follows. Its $(2,1)$ superspace formulation is given in Section 3  and general $(4,1)$ sigma model actions written in $(2,1)$ superspace are introduced in Section  4 and the geometric conditions for $(4,1)$ supersymmetry are studied.
 Related $(4,2)$ multiplets are discussed in Section 5. General 
  $(4,2)$ supersymmetric sigma models are studied in $(2,2)$ superspace in Section 6 and the conditions for $(4,2)$ supersymmetry are analysed. The relationship to the $(4,4)$ hypermultiplet is discussed in Section 7, while Section 8 contains results on the $(4,1)$ superspace action.  In Section 9 we introduce $(4,q)$ projective superspace and use it to formulate multiplets and actions, giving explicit constructions of target space geometries.

 \section{$(4,1)$ Off-Shell Supermultiplets}
\label{Off}

In \cite{Gates:1984nk}, a $(4,4)$ off-shell multiplet was 
found by dimensional reduction of $N=2$ super Yang-Mills theory from 4 dimensions.
Truncating this  gives an off-shell $(4,1)$ supermultiplet that can be formulated as follows. We use
 $(4,1)$ superspace with coordinates $x^\+,x^=, \theta ^+_a, \bar \theta ^{+a},\theta^-$  where the index $a=1,2$ is an $SU(2)$ index. Here $\theta ^+_a$ are complex and $\theta^-$ is real.
 There are
 two right-handed complex spinorial covariant derivatives $\bbD{+}^a$ and a real 
 left-handed spinorial covariant derivative $D_-$, satisfying
\ber\nn\label{talg1}
&&\{\bbD{+a},\bbDB{+} ^b\}=~2i\delta^b_a\pa_\+~, ~~~a,b,=1,2.\\[1mm]
&&(D_-)^2=i\pa_=~.
\eer
 The $(4,1)$ multiplet  obtained  by truncating the  $(4,4)$ multiplet of  \cite{Gates:1984nk}
 consists of a pair of $(4,1)$ superfields $\phi, \chi$ satisfying the constraints 
\ber\nn\label{constr2}
&&\bbDB{+}^1\phi = 0=\bbD{+2}\phi~,~~~\bbDB{+}^1 \chi =0=\bbD{+2}\chi~,\\[1mm]
&&\bbDB{+}^2\chi=-i\bbDB{+}^1\bar \phi~,~~~\bbDB{+}^2\phi=i\bbDB{+}^1\bar\chi~.
\eer

The supersymmetry transformations can be put into the form \re{nis1tfs} by expanding in  $(1,1)$  superspace.
The $(4,1)$ multiplet in \re{constr2} can be formulated in $(1,1)$   superspace
by defining
\ber\label{1comp}
\phi \big\vert _{\theta _2^+=0,\theta _1^+=\bar \theta _1^+} = \tilde \phi , \qquad \chi \big \vert _{\theta _2^+=0,\theta _1^+=\bar \theta _1^+} = \tilde \chi~.
\eer
The constraints \re{constr2} then determine the terms in $\phi, \chi $ of higher order in $\theta_2^+,\theta _1^+-\bar \theta _1^+$
in terms of $\tilde \phi, \tilde \chi $
and give the supersymmetry transformations under the non-manifest supersymmetries.
We define four real $(4,1)$ superspace spinor derivatives $D_+$ and $\check{D}_+^{(A)}~,~A=1,2,3$ by
\ber\nn\label{8}
&&\bbD{+1}=: D_+ - i\check{D}_+^{(1)}\\[1mm]
&&\bbD{+2}=: \check{D}_+^{(2)} - \check{D}_+^{(3)}~.
\eer
Then  $D_+$ is 
the $(1,1)$ superspace spinor derivative  and the three differential operators $\check{D}_+^{(A)}$, $A=1,2,3$, determine the generators of nonmanifest supersymmetries $Q_+^{(A)}$ via the constraint \re{constr2} 
\ber
\check{D}^{(A)}_+\phi\Big\vert _{\theta _1^+ = \bar \theta _1^+,\theta _2^+=0}& =&Q^{(A)}_+ \tilde \phi~,
\\
\check{D}^{(A)}_+\chi\Big\vert _{\theta _1^+ = \bar \theta _1^+,\theta _2^+=0}& =&Q^{(A)}_+ \tilde \chi~.
\eer
This results in the following relation for the extended supersymmetries for $d$ superfields,
\ber
Q_+^{(A)}\left(\begin{array}{c}\tilde\phi\\
\tilde\chi\\
\bar{\tilde\phi}\\
\bar{\tilde\chi}\end{array}\right)=:\mathbb{J}^{(A)}D_+\left(\begin{array}{c}\tilde\phi\\
\tilde\chi\\
\bar{\tilde\phi}\\
\bar{\tilde\chi}\end{array}\right)~,\eer
where the complex structures
\ber\label{comstr}
\mathbb{J}^{(A)}= \mathbb{I}^{(A)}\otimes \mathbb{1}_{d\times d}
\eer
with
\ber\label{comstr1}
\mathbb{I}^{(1)}=\left(\begin{array}{cc}i\mathbb{1}&0\\
0&-i\mathbb{1}\end{array}\right)~,~~~
\mathbb{I}^{(2)}=\left(\begin{array}{cc}0& i\sigma_2\\ i\sigma_2&0\end{array}\right)
~,~~~
\mathbb{I}^{(3)}=\left(\begin{array}{cc}0&-\sigma_2\\
\sigma_2&0\end{array}\right)
\eer
 are constant in this coordinate system and satisfy the quaternion algebra
\ber
\mathbb{J}^{(A)}\mathbb{J}^{(B)}=-\delta^{AB}+\epsilon^{ABC}\mathbb{J}^{(C)}~.
\eer
Then this gives transformations for $\tilde \phi, \tilde \chi $ of the form \re{nis1tfs}.

\section{$(2,1)$  Superspace Formulation}

The general  $(2,1)$ sigma model action   can be written in $(2,1)$  superspace as  \cite{Hull:1985jv,AbouZeid:1997cw}
\ber\label{2,1}
S= 
\int d^2x d^3 \theta
\left(k_ \alpha D_-\varphi^\alpha
+
\bar k_{\bar \alpha}  D_-\bar \varphi^{\bar \alpha} 
\right)~.
\eer
The fields  $\varphi^\alpha$ are $(2,1)$ chiral 
\ber
\bbDB{+}\varphi^\alpha=0~,
\eer
and $\bar \varphi^{\bar \alpha} $ are their complex conjugates $\bar \varphi^{\bar \alpha} =
(\varphi^\alpha)^*$.
The 
theory is defined locally by a 1-form potential $k_\alpha (\varphi, \bar  \varphi)$ with 
$
\bar k_{\bar \alpha} = (k_\alpha)^*$, which is
 defined up to the addition of the gradient of a function $h(\varphi,\bar\varphi) $ and a holomorphic 1-form $f_\alpha (\varphi)$, 
 \ber\label{kfreedom}
 k_\alpha(\varphi,\bar\varphi) \to  k_\alpha(\varphi,\bar\varphi)+\pa_\alpha h(\varphi,\bar\varphi) +  f_\alpha (\varphi)~.
\eer
The metric $g$ and $B$ field  for the model \re{2,1} are (in a particular gauge)\cite{AbouZeid:1997cw}
\ber\nn\label{gB}
&&g_{\alpha \bar \beta}=i(\pa_{ \alpha} \bar k_{\bar  \beta}-\pa_{\bar \beta} k_{\alpha} )           \\[1mm]\nn
&&B^{(2,0)}_{\alpha\beta}=  i(  
\pa_{\alpha} k_{\beta}     -\pa_{  \beta} k_{\alpha}   )
\\[1mm]
&&B=B^{(2,0)}+B^{(0,2)}
\eer
as may be verified by reducing to  the $(1,1)$ superspace formulation \cite{Hull:1985jv,AbouZeid:1997cw}.

The $(4,1)$ multiplet \re{constr2} can be expanded into $(2,1)$  superspace
by writing\footnote{We temporarily use the tilde notation for the $(2,1)$ components in this section, just as  we did  for the $(1,1)$ components in section \ref{Off}.}
\ber
\phi \vert _{\theta _2^+=0} = \tilde \phi , \qquad \chi \vert _{\theta _2^+=0} = \tilde \chi~.
\eer
The constraints \re{constr2} then define the terms in $\phi, \chi $ of higher order in $\theta_2$
and give the supersymmetry transformations under the non-manifest supersymmetries.
The $(4,1)$ derivative $\bbD{+1}$ survives as  the  (2,1) derivative $\bbD{+}$
{while 
$\bbD{+2}$ gives the generator $Q$ of non-manifest supersymmetries, acting as:
\ber\label{quis}
\bar Q_+\tilde \phi=(\bbDB{+}^2 \phi )\Big| _{\theta^2=0} \, , \qquad \bar Q_+ \tilde \chi=(\bbDB{+}^2 \chi )\Big|_{\theta^2=0}
\, , \qquad \bar Q_+ \bar  {\tilde \phi }=0
\, , \qquad \bar Q_+ \bar  {\tilde \chi }=0
~.
\eer
Complex conjugation then gives  the action of the  generator $Q$.

The action for $d$ $(4,1)$  multiplets must take the form \re{2,1} when written in $(2,1)$  superspace, with
$(2,1)$  chiral superfields  $\varphi^\alpha = (\tilde \phi ^i, \tilde \chi ^i)$ with $i=1,\dots , d$.
We will henceforth drop the tildes on $\tilde \phi ^i, \tilde \chi ^i$.
Then using the constraints \re{constr2}, \re{quis} gives the non-manifest supersymmetry transformations 
\ber\label{spec}
\bar Q_+ \phi= i\bbDB{+} \bar  {\chi }, \qquad \bar Q_+ \chi
= -i\bbDB{+} \bar  {\phi } \, , \qquad \bar Q_+ \bar  {  \phi }=0
\, , \qquad \bar Q_+ \bar  {  \chi }=0~.
\eer }
The potential has components
$k_\alpha = (k_{\phi^i},k_{\chi^i})$ 
and the variation of the action \re{2,1} under the non-manifest supersymmetries generated by 
$\bar Q_+$ takes the form
\ber
\label{erterwf}
\delta S=\int d^2x \bbD{+}\bbDB{+}D_- \Delta
\eer
where 
\ber\nn
\Delta=
&&iD_-\phi^i\left(k_{\phi^i ,\chi^j}\bbDB{+}\bar\phi^j-k_{\phi^i, \phi^j}\bbDB{+}\bar\chi^j\right)\\[1mm]\nn
&&-iD_-\bar\phi^i\left[(\bar k_{\bar\phi^i, \phi^j}+k_{\chi^i\bar\chi^j})\bbDB{+}\bar\chi^j-(\bar k_{\bar\phi^i ,\chi^j}-k_{\chi^i , \bar\phi^j})\bbDB{+}\bar\phi^j\right]
\\[1mm]
&&-(\phi\leftrightarrow\chi)~.
\eer
The  second line vanishes if $k$ satisfies 
\ber\nn\label{vpotcs}
&&k_{\phi^i,\bar \phi^{ j}}+\bar k_{\bar\chi^i, \chi^{ j}}=0~,\\[1mm]
&&k_{\phi^i,\bar \chi^ j}-\bar k_{\bar\chi^i, \phi^j}=0~,
\eer
where  the comma denotes  a partial derivative, so that e.g. $k_{\phi^i,\bar \phi^{ j}}= \pa
k_{\phi^i} / \pa \bar \phi^{ j}
$.
Then $\bbDB{+}\Delta $
gives expressions that vanish after repeated use of  \re{vpotcs} and their derivatives.
Thus
\re{vpotcs} implies that the variation (\ref{erterwf}) of the action under the extra supersymmetries vanishes.

We note that the vanishing of $\Delta$ and $\bbDB{+}\Delta $ is sufficient for invariance, but not   necessary. For invariance, it is only necessary that they  reduce to terms that vanish when integrated,
so that $\bbD{+}\bbDB{+}D_- \Delta $ is a total derivative with 
$\int d^2x \bbD{+}\bbDB{+}D_- \Delta=0$ (up to a boundary term).
This is essentially the condition that the variation of the action under the non-manifest supersymmetries can be cancelled by transformations of the form
 \re{kfreedom}.
The  full necessary and sufficient conditions for supersymmetry 
will be given in the next section, from a  
 geometric analysis.
 We will return to the $(4,1)$ superspace formulation of these actions 
in sections \ref{general} and \ref{GEN}.

\section{General $(4,1)$ Sigma Models}
\label{general}

We now consider the general conditions for the $(2,1)$  superspace  action (\ref{2,1}) to  be $(4,1)$ supersymmetric so that it is invariant under two further supersymmetries.
Following \cite{Hull:1985pq} and  \cite{Goteman:2009ye}, we make the ansatz
\ber\nn
\label{fvar}
&&\delta\varphi^\alpha =\bar\epsilon^+\bbDB{+}f^\alpha(\varphi,\bar\varphi)\\[1mm]
&&\delta\bar\varphi^{\bar\alpha} = \epsilon^+\bbD{+}\bar f^{\bar\alpha}(\varphi,\bar\varphi)
\eer
for the additional  supersymmetries of the action \re{2,1}.  Up to central charge transformations, this is the most general ansatz compatible with the chirality properties \cite{Goteman:2009ye}.

Expanding  in components and comparing with (\ref{nis1tfs}), we can read off the form of the complex structures.
The manifest (2,1) supersymmetry involves
 the canonical complex structure
\ber
\mathbb{J}^{(1)}=\left(\begin{array}{cc}i\mathbb{1}&0\\
0&-i\mathbb{1}\end{array}\right)\, ,
\eer
while the  transformation (\ref{fvar}) yields  second and third ones 
\ber\label{compl2}
\mathbb{J}^{(2)}=\left(\begin{array}{cc}0&f^\alpha_{~\bar\beta}\\\bar f^{\bar\alpha}_{~\beta}&0\end{array}\right)~,~~~\mathbb{J}^{(3)}=\left(\begin{array}{cc}0&if^\alpha_{~\bar\beta}\\-i\bar f^{\bar\alpha}_{~\beta}&0\end{array}\right)~.
\eer
Here, the lower index on $f$ denotes a derivative,
\ber
f^\alpha _{~\bar\beta}:=\frac {\partial f^\alpha} {\partial \bar\varphi^{\bar\beta}}~.
\eer

From off-shell closure of the algebra, 
\ber\nn
[\delta_1,\delta_2]\varphi^\alpha=2i\epsilon^+_{[2}\bar\epsilon^+_{1]}\partial_\+\varphi^\alpha _,
\eer
we deduce that (c.f.\cite{Hull:1985pq})
\ber\nn
&&f^\alpha_{~\bar\beta}\bar f^{\bar\gamma}_{~\alpha}=-\delta^{\bar\gamma}_{~\bar\beta}~,~~~\bar f^{\bar\alpha}_{~\beta}f^{\gamma}_{~\bar\alpha}=-\delta^{\gamma}_{~\beta}~,\\[1mm]
&& f^\alpha_{~[\bar\alpha}f^\beta_{~\bar\beta]\alpha}=0~,~~~ \bar f^{\bar\alpha}_{~[\alpha}\bar f^{\bar\beta}_{~\beta]\bar\alpha}=0~.
\eer
(See \cite{Hull:1985pq} 
for similar relations for $N=2$ in $d=4$.)
Here $f^\beta_{~\bar\beta\alpha}= 
\partial f^\beta/ \partial \bar\varphi^{\bar\beta}\partial  \varphi^\alpha$ etc.
Then the matrices   $\mathbb{J}^{(1)},\mathbb{J}^{(2)},\mathbb{J}^{(3)}$
satisfy the quaternion algebra and have vanishing Nijenhuis tensors 
\ber\label{nij}
{\cal N}^i_{jk}(\mathbb{J}^{(A)})=0~,
\eer
 so that they are each complex structures.

 The remaining geometric constraints follow from invariance of the action.
Varying the action we find 
\ber
\delta S=  \int d^2x d^3 \theta \,  \Delta
\eer
where
\ber\label{varS}
 \Delta=
\bar\epsilon^+\bbDB{+}f^\beta\left(B_{\beta\alpha}D_-\varphi^\alpha+g_{\beta\bar\alpha}D_-\bar\varphi^{\bar\alpha}\right)~.
\eer
Pushing in $\bbDB{+}$ from the measure yields\footnote{If we performed the full reduction to  $(1,1)$ this would to parallel  the calculation in \cite{Gates:1984nk} for $(2,2)$ supersymmetry.}
\ber\label{mess}
\bbDB{+} \Delta=
\bar\epsilon^+\bbDB{+}f^\beta\bbDB{+}\bar\varphi^{\bar\beta}\left(B_{\beta\alpha ,\bar\beta}D_-\varphi^\alpha+g_{\beta\bar\alpha, \bar\beta}D_-\bar\varphi^{\bar\alpha}\right)-\bar\epsilon^+\bbDB{+}f^\beta D_-\bbDB{+}\bar\varphi^{\bar\alpha}g_{\beta\bar\alpha}~.
\eer
Integrating $D_-$ by parts and defining 
\ber
\omega_{\bar\beta\bar\alpha}:=f^\beta_{~[\bar\alpha}g_{\bar\beta]\beta}
=
\frac 1 2 \left(
f^\beta_{~\bar\alpha}g_{\beta\bar\beta} -f^\beta_{~\bar
\beta
}g_{\beta\bar \alpha
}
\right)
\eer we rewrite the last term as
\ber\label{messi}
-\bar\epsilon^+\left(\bbDB{+}\bar\varphi^{\bar\beta}D_-\bbDB{+}\bar\varphi^{\bar\alpha}f^\beta_{~(\bar\beta}g_{\beta\bar\alpha)}+\half D_-\omega_{\bar\beta\bar\alpha}\bbDB{+}\bar\varphi^{\bar\alpha}\bbDB{+}\bar\varphi^{\bar\beta}\right)
-\half D_- \Bigl( \epsilon^+\omega_{\bar\beta\bar\alpha}\bbDB{+}\bar\varphi^{\bar\alpha}\bbDB{+}\bar\varphi^{\bar\beta}
\Bigr)
\eer
and drop the final term here as it is a total derivative.

Then the condition for supersymmetry is
\ber\label{messy}\nn
&&\bar\epsilon^+\bbDB{+}f^\beta\bbDB{+}\bar\varphi^{\bar\beta}\left(B_{\beta\alpha ,\bar\beta}D_-\varphi^\alpha+g_{\beta\bar\alpha, \bar\beta}D_-\bar\varphi^{\bar\alpha}\right)
\\
&&-\bar\epsilon^+\left(\bbDB{+}\bar\varphi^{\bar\beta}D_-\bbDB{+}\bar\varphi^{\bar\alpha}f^\beta_{~(\bar\beta}g_{\beta\bar\alpha)}+\half D_-\omega_{\bar\beta\bar\alpha}\bbDB{+}\bar\varphi^{\bar\alpha}\bbDB{+}\bar\varphi^{\bar\beta}\right)=0~.
\eer
Then the independent terms in \re{messy} give the equations \footnote{{Pushing in additional $\bbD{}$s from the measure and/or partial integration of bosonic derivatives does not relate these terms or lead to any further simplifications.}}  
\ber\label{geom}
&&f^\beta_{~(\bar \alpha}g_{\bar\gamma)\beta}=0~,~~~\Rightarrow f^\beta_{~\bar \alpha}g_{\bar\gamma\beta}=
\omega_{\bar \gamma\bar\alpha}~,
\eer
together with
\ber
\label{geom2}
\nn
&& \half \omega_{\bar\alpha\bar\gamma,\bar\beta}-g_{\beta\bar\beta,[\bar\alpha}
 f^\beta_{~\bar\gamma]}=0~,~~~\Rightarrow \na^{(+)}_{\bar\beta}\omega_{\bar\alpha\bar\gamma}=0~,\\[1mm]
&&\half\omega_{\bar\alpha\bar\gamma,\beta}-B^{(2,0)}_{\sigma\beta ,[\bar\alpha}f^{\sigma}_{~\bar\gamma]}=0~ ~,~~~\Rightarrow \na^{(+)}_{\beta}\omega_{\bar\alpha\bar\gamma}=0~,
\eer
where we have used the geometric constraints on the connection and torsion that follow from the underlying $(2,1)$ geometry, as well as the definitions \re{gB}. Some of this structure is described in Appendix \ref{app1}. 
The conditions  \re{geom} imply that the metric is hermitian with respect to the complex structures \re{compl2} while \re{geom2} implies
 that these complex structures are covariantly constant with respect to the  
 connection with torsion $\Gamma^{(+)}=\Gamma^{(0)}+T$: 
\ber \label{nabf}
\na^{(+)}_i f^\kappa{}_{\bar\lambda}=0~, 
\eer
where $\Gamma^{(0)}$ is the levi-Civita connection and the torsion is formed from the $B$  field strength as $T=\half g^{-1}H$. 
We note that this geometry   is sometimes referred to as hyperk\"ahler with torsion.
{Finally,  the vanishing of the Nijenhuis tensor  \re{nij} in conjunction with the covariant constancy conditions in \re{geom2} leads to
\ber
H=
d^{(A)} \omega^{(A)}~,
\eer
for each $A$, where $\omega^{(A)}$ is the 2-form with components $\omega^{(A)}_{ij} =g_{ik} (\mathbb{J}^{(A)})^k{}_j$, and   $d^{(A)}$ is the $i(\bar\pa -\pa)$ operator for the   complex structure $\mathbb{J}^{(A)}$. This can also be  derived from $\na^{(+)}J^{(A)}=0$ and ${\cal N}^{~~i}_{jk}(J^{(A)})=0$.}

The transformations \re{fvar} correspond to generalising  the constraints \re{constr2} to
\ber\nn\label{constr22}
&&\bbDB{+}^1\varphi^\alpha  = 0 ~,\\[1mm]
&&\bbDB{+}^2\varphi^\alpha = f^\alpha  _{~\bar\beta}\bbDB{+}^1\bar \varphi ^{\bar\beta}
\eer
in $(4,1) $ superspace.
Note that the constraints \re{fvar} require the existence of a local product structure in addition to the structure required for $(4,1)$ geometry, as this is necessary to split the coordinates into two sets,  $\varphi = (\phi, \chi)$. For $(4,2)$ or $(4,4)$ supersymmetry, the existence of this product structure follows from the conditions for extended supersymmetry.

For constant complex structures $f^\alpha_{~\bar\beta}$, \re{fvar}  implies that
\ber
\Gamma^{(+)}_{i\sigma\bar\kappa}f^\sigma{}_{\bar\lambda}+ \Gamma^{(+)}_{i\bar\lambda\sigma}f^\sigma{}_{\bar\kappa}=0~,
\eer
where $i=(\beta,\bar\beta)$,   we have lowered $\kappa$ to $\bar\kappa$ and used the antisymmetry of the two-forms $\omega$.   
This (non-covariant) condition can be rewritten using formulae from the appendix as
\ber\nn\label{noncov}
&&f^\sigma_{~\bar\lambda}g_{\sigma\bar\kappa,\bar\beta}+2g_{\sigma \bar\beta,[\bar\lambda}f^\sigma_{~\bar\kappa]}=0\\[1mm]
&&f^\sigma_{~\bar\kappa}g_{\bar \lambda\sigma,\beta}+2f^\sigma_{~[\bar\lambda}g_{\bar\kappa]\beta,\sigma}=0~.
\eer
For the constant complex structures \re{comstr},
we have
\ber
f^\alpha_{~\bar\beta}= i( \sigma _2) ^\alpha_{~\bar\beta}
\eer
and
 the  hermiticity condition  \re{geom} becomes
\ber\label{herm1}\nn
&&\bar k_{\bar\phi^i,\phi^j}-k_{\phi^j,\bar\phi^i}-\bar k_{\bar\chi^j,\chi^i}+k_{\chi^i,\bar\chi^j}=0\\[1mm]
&&\bar k_{\bar\chi^{(i},\phi^{j)}}-k_{\phi^{(i},\bar\chi^{j)}}=0~,\eer
while the covariant constancy conditions 
\re{geom2} or \re{noncov}
become
\ber\nn\label{herm2}
&&\half\left(k_{\phi^{[j} ,\bar\chi^{k]}}-\bar k _{\bar\chi^{[j},\phi^{k]}}\right)_{, \bar\beta}-\bar k_{\bar\beta, \phi^{[j} \bar\chi^{k]}}=0\\[1mm]\nn
&&\half\left(\bar k_{\bar\phi^k,\phi^j}+ k_{\chi^k, \bar\chi^j}+\bar k_{\bar\chi^j,\chi^k}+ k_{\phi^j, \bar\phi^k}\right)_{, \bar\beta}
-\bar k_{\bar\beta, \phi^j\bar  \phi^k}-\bar k_{\bar\beta, \chi^k \bar\chi^j}=0\\[1mm]
&&\half\left(k_{\chi^{[j}, \bar\phi^{k]}}-\bar k _{\bar\phi^{[j}, \chi^{k]}}\right)_{,
\bar\beta}-\bar k_{\bar\beta, \chi^{[j}
\bar\phi^{k]}}=0~.
\eer
We note that  if \re{vpotcs} are satisfied, then this implies that \re{herm1} and \re{herm2} are satisfied. The converse is not true, and  \re{vpotcs} gives a special case of the general  conditions \re{herm1} and \re{herm2}. E.g. \re{vpotcs} requires that $k_{\phi^i,\bar \phi^{ j}}+\bar k_{\bar\chi^i, \chi^{ j}}$ vanishes whereas \re{herm1}  only sets it equal to its hermitean conjugate.

\section{$(4,2)$ Off-Shell Supermultiplets}

Truncating the 
$(4,4)$ off-shell multiplet of \cite{Gates:1984nk} to $(4,2)$ superspace gives
 an off-shell $(4,2)$ supermultiplet that can be formulated as follows. We use
 $(4,2)$ superspace with coordinates $x^\+,x^=, \theta ^{+a}, \bar \theta ^{+}_a,\theta^-, \bar \theta^-$  where  $a=1,2$ is an $SU(2)$ index.\footnote{ There is a possible confusion between the $SU(2)$ index $2$ and a $2$  indicating the square. This is resolved  by noting that a bold face $\bbD{\pm} $ never appears squared.}   All fermionic coordinates are complex.
 There are
 two complex right-handed spinorial covariant derivatives $\bbD{+}^a$ and a complex 
 left-handed spinorial covariant derivative $\bbD{-}$, satisfying
\ber\nn\label{talg}
&&\{\bbD{+a},\bbDB{+} ^b\}=~2i\delta^b_a\pa_\+~, ~~~a,b,=1,2,\\[1mm]
&&\{\bbD{-}, \bbDB{-}\}= 2i\pa_=~,
\eer

 The $(4,2)$ multiplet obtained   from truncating the  $(4,4)$ multiplet of  \cite{Gates:1984nk}
 consists of a pair of  $(4,2)$ superfields $\phi, \chi$ satisfying the constraints 
\ber\nn\label{constr21}
&&\bbDB{+}^1\phi = 0=\bbD{+2}\phi~,~~~\bbDB{+}^1 \chi =0=\bbD{+2}\chi~,\\[1mm]\nn
&&\bbDB{+}^2\chi=-i\bbDB{+}^1\bar \phi~,~~~\bbDB{+}^2\phi=i\bbDB{+}^1\bar\chi,\\[1mm]
&&\bbDB{-} \phi =0 ~,~~~ \bbD{-} \chi =0. 
\eer

An alternative truncation has  the $\bbD{-}$ constraints on the two fields switched. The two multiplets are related by interchanging $ \theta_- \leftrightarrow \bar \theta _-$, so a theory written in terms of one multiplet is equivalent to one written in terms of the other. Indeed, we show in section \ref{4.2} that their projective superspace formulations are isomorphic.
However, just as for the $(2,2)$ chiral and twisted chiral multiplets, one might suspect that  there could be new non-trivial theories that have both kinds of supermultiplet. As far as we have been able to ascertain, this is not the case (as long as no further superfields are involved) as no supersymmetric interaction between the two kinds of multiplets seems possible\rd{\footnote{Added in proof: The referee informs us that this is in agreement with the results of \cite{Ivanov:2004re}  derived using bi-harmonic superspace}}.

\section{$(4,2)$ Supersymmetry in $(2,2)$ Superspace}

In $(2,2)$ superspace, chiral superfields
$\varphi$
satisfy
\ber
\bbDB{\pm} \varphi =0
\eer
while twisted chiral superfields $\psi$ satisfy
\ber
\bbDB{+} \psi =0, \qquad \bbD{-}\psi=0
\eer
There are other possible $(2,2)$ multiplets such as semichiral multiplets \cite{Buscher:1987uw}, but here we shall restrict ourselves to 
these two.

The general action for chiral and twisted chiral multiplets is given 
by \cite{Gates:1984nk}
\ber
S= \int d^2x d^4\theta \,K ( \varphi, \bar \varphi, \psi, \bar \psi ) 
\eer
in terms of an unconstrained scalar potential $K( \varphi, \bar \varphi, \psi, \bar \psi ) $.
Expanding in $(2,1)$  superfields
by writing
\ber
\varphi \vert _{\theta _2^-=0} = \tilde \varphi , \qquad \psi \vert _{\theta _2^-=0} = \tilde \psi
\eer
one finds the action
\re{2,1} and the
vector potentials are gradients of the scalar potential $K$ \cite{Hull:2012dy},
\ber
k_{\varphi}= i \partial_{\varphi}K~,~~~k_{\psi}=-i \partial_{\psi}K.
\label{kKcom}
\eer
where the tildes and indices enumerating multiplets have been suppressed.

We now turn to the off-shell $(4,2)$ supermultiplet introduced in the last section.
It contains a $(2,2)$ chiral superfield $\phi$ and a twisted chiral superfield $\chi$ with the transformation under the {extra supersymmetries $Q, \bar Q$ given by 
\ber
\bar Q_+ \phi = i\bbDB{+} \bar \chi, \qquad \bar Q_+ \chi = -i\bbDB{+} \bar \phi
, \qquad \bar Q_+ \bar \phi=0
, \qquad \bar Q_+ \bar \chi=0
~,
\eer
together with the complex conjugate expressions.}

Consider  a model with $d$ multiplets $\phi^i, \chi^i$, so the action is
\ber
S= \int d^2x d^4\theta \,K ( \phi^i, \chi^i,\bar\phi^i, \bar\chi^i ) ~.
\eer
Then under the 
$\bar Q$ transformation
\ber
\delta S=\int d^2x \bbD{+}\bbDB{+}\bbD{-}\bbDB{-} \Delta
\eer
where 
\ber
\Delta = \bar QK =i  K,_{\phi ^i} \bbDB{+} \bar \chi ^i -iK,_{\chi ^i} \bbDB{+} \bar \phi ^i ~.
\eer
Then acting with $\bbDB{+}$ gives
\ber
\delta S=\int d^2x \bbD{+}\bbD{-}\bbDB{-} (\bbDB{+} \Delta)
\eer
where 
\ber
\bbDB{+}\Delta = \bbDB{+}\bar QK =\bbDB{+}(i  K,_{\phi ^i} \bbDB{+} \bar \chi ^i -iK,_{\chi ^i} \bbDB{+} \bar \phi ^i )~.
\eer
This gives 
\ber\nn
\bbDB{+}\Delta =
&&
i  K,_{\phi ^i \bar \phi ^j} \bbDB{+} \bar \phi ^j \bbDB{+} \bar \chi ^i 
+i  K,_{\phi ^i   \bar \chi ^j }    \bbDB{+} \bar \chi ^j \bbDB{+} \bar \chi ^i 
\\ &&-iK,_{\chi ^i \bar \phi ^j}  \bbDB{+} \bar \phi ^j\bbDB{+} \bar \phi ^i 
-iK,_{\chi ^i \bar \chi ^j } \bbDB{+} \bar \chi ^j  \bbDB{+} \bar \phi ^i 
\\
 \nn
=
&&
i  
(K,_{\phi ^i \bar \phi^j} + K,_{\chi ^j \bar \chi^i})
 \bbDB{+} \bar \phi ^j \bbDB{+} \bar \chi ^i 
\\ &&+i  K,_{\phi ^i   \bar \chi ^j }    \bbDB{+} \bar \chi ^j \bbDB{+} \bar \chi ^i 
-iK,_{\chi ^i \bar \phi ^j}  \bbDB{+} \bar \phi ^j\bbDB{+} \bar \phi ^i .
\eer
The first term vanishes if 
\ber\label{4.1cnd}
 K,_{\phi ^i \bar \phi^j} + K,_{\chi ^j \bar \chi^i} =0~.
\eer
This is a sufficient condition for full invariance, since using it one finds that the remaining terms vanish using $\bbDB{-} $ or $\bbD{-} $ from the remaining measure:
\ber\nn
&&\bbDB{-} (K,_{\phi ^i   \bar \chi ^j }    \bbDB{+} \bar \chi ^j \bbDB{+} \bar \chi ^i)=0\\[1mm]
&&\bbD{-} (K,_{\chi ^i \bar \phi ^j}  \bbDB{+} \bar \phi ^j\bbDB{+} \bar \phi ^i )=0~.
\eer

To find the necessary and sufficient conditions for $(4,2)$ supersymmetry, we
start with the conditions for $(4,1)$ supersymmetry given by \re{herm1} and \re{herm2}.
For the sigma model to have $(4,2)$ supersymmetry requires in addition the condition (\ref{kKcom}) which here implies that the $(4,1)$ potential $k$ is given by derivatives of a scalar potential $K$:
\ber\label{VecScal}
k_{\phi ^i}=iK,_{\phi ^i}, ~~~k_{\chi ^i} =-iK,_{\chi ^i} ~.
\eer 
Then the hermiticity condition \re{herm1} together with \re{VecScal} gives precisely the condition \re{4.1cnd}, and then the remaining conditions \re{herm2} are all satisfied identically using
 \re{4.1cnd} and  \re{VecScal}, and  give no further constraints.
 Thus \re{4.1cnd} is the necessary and sufficient condition for a $(2,2)$ model to have $(4,2)$ supersymmetry.

In section 3, we considered $(4,1)$ models whose potentials satisfied the conditions (\ref{vpotcs}). These models will have $(4,2)$ supersymmetry if  \re{VecScal} is satisfied, which implies \re{4.1cnd} together with 
\ber
K,_{\phi ^i \bar \chi^j} =K,_{\phi ^j \bar \chi^i} ~.
\eer
This gives a special class of $(4,2)$ models.

\section{$(4,4)$ Supermultiplet and Action}
\label{4,4}
The 
$(4,4)$ off-shell multiplet of \cite{Gates:1984nk} is formulated in (4,4) superspace
with
 two complex right-handed spinorial covariant derivatives $\bbD{+a}$ and two complex 
 left-handed spinorial covariant derivatives $\bbD{-a}$, satisfying
\ber\nn\label{talg44}
&&\{\bbD{+a},\bbDB{+} ^b\}=~2i \delta^b_a\pa_\+~, ~~~a,b,=1,2.\\[1mm]
&&\{\bbD{-a}, \bbDB{-}^b\}= 2i \delta^b_a \pa_=~,
\eer
 The  $(4,4)$ multiplet of  \cite{Gates:1984nk}
 consists of a pair of superfields $\phi, \chi$ satisfying the constraints 
\ber\nn\label{constr44}
&&\bbDB{+}^1\phi = 0=\bbD{+2}\phi~,~~~\bbDB{+}^1 \chi =0=\bbD{+2}\chi~, ~,~~~ \bbD{-a} \chi =0\\[1mm]\nn
&&\bbDB{+}^2\chi=-i\bbDB{+}^1\bar \phi~,~~~\bbDB{+}^2\phi=i\bbDB{+}^1\bar\chi,\\[1mm]
&&\bbDB{-}^2\chi= i\bbD{-1} \phi~,~~~\bbD{-2}\phi=i\bbDB{-}^1\chi~.
\eer

As before, the action can be written in (2,2) superspace in terms of
 $d$ (2,2) chiral multiplets $\phi^i$ and $d$ twisted chiral multiplets $\chi^i$, so the action is
\ber
S= \int d^2x d^4\theta \,K ( \phi^i, \chi^i,\bar\phi^i, \bar\chi^i ) 
\eer
with the non-manifest supersymmetry transformations given by
{
\ber
\bar Q_+ \phi = i\bbDB{+} \bar \chi, \qquad \bar Q_+ \chi = -i\bbDB{+} \bar \phi
, \qquad \bar Q_+ \bar \phi=0
, \qquad \bar Q_+ \bar \chi=0
~,
\eer
and
\ber
Q_- \phi = i \bbDB{-} \chi, \qquad Q_- \bar \chi = -i    \bbDB{-} \bar \phi
, \qquad Q_-\bar \phi=0
, \qquad Q_- \chi=0
~,
\eer
together with the complex conjugate expressions.}

Then under the  
$\bar Q_+$ transformation
\ber
\label{vardel}
\delta S=\int d^2x \bbD{+}\bbDB{+}\bbD{-}\bbDB{-} \Delta
\eer
where  
\ber
\Delta = \bar Q_+K =i  K,_{\phi ^i} \bbDB{+} \bar \chi ^i -iK,_{\chi ^i} \bbDB{+} \bar \phi ^i 
\eer
and, as in the last section, the action is invariant if
\ber
\label{erty}
 K,_{\phi ^i \bar \phi^j} + K,_{\chi ^j \bar \chi^i} =0~.
\eer
Under the $Q_-$ transformation we obtain (\ref{vardel}) but with
\ber
\Delta = Q_-K =i  K,_{\phi ^i}
 \bbDB{-} \chi ^i
  -iK,_{\bar \chi ^i} 
\bbDB{-} \bar \phi  
  ^i 
\eer
Then a similar analysis to the above gives that the action is invariant under the $Q_-$ transformation if
\ber
\label{ertyb}
 K,_{\phi ^i \bar \phi^j} + K,_{\chi ^i \bar \chi^j} =0~.
\eer
Then the necessary and sufficient conditions for $(4,4)$ supersymmetry are  (\ref{erty}) and
 (\ref{ertyb}).

Together, (\ref{erty}) and
 (\ref{ertyb}) 
imply
\ber
\label{ertyc}
 K,_{\phi ^i \bar \phi^j} =K,_{\phi ^j \bar \phi^i}~.
\eer
We can then instead take the necessary and sufficient conditions for $(4,4)$ supersymmetry to be (\ref{ertyc}) and
 (\ref{ertyb}), which are precisely
 the conditions that were found in  \cite{Gates:1984nk}.

\section{$(4,1)$ Superspace Action}
\label{GEN}
\subsection{General}
\label{gen}
A superspace action  for ${\cal N}$ supersymmetries in $D$ dimensions
involves integration over the $d=s{\cal N}$ fermi coordinates $\theta$, where $s$ is the dimension of the spinor representation in $D$ dimensions
(e.g. $s=4$ in $D=4$).
This picks out the highest $\theta$ component from the superspace Lagrangian ${\cal L}$. Equivalence between Berezin integration and differentiation means that the integration may be  written schematically as
\ber
\int d^Dx d^d\theta{\cal L}=\int d^Dx \frac {\partial^d{\cal L}}{\partial\theta^d}=\int d^Dx D^d{\cal L}\Big| _{\theta =0}~,
\eer
where the vertical bar denotes the $\theta$-independent part of the expression and use has been made of the fact that the spinorial covariant derivatives $D$ differ from the partial spinorial derivatives by $\theta$ terms involving  a spacetime derivative, and  total derivative terms are dropped from the spacetime integral. Since the product $DD\sim\partial$, with $\pa$ a space time derivative, it is clear that 
even if the  Lagrangian ${\cal L}$ contains no derivatives, there is a limit to $d \leq 4$ in spacetime dimensions $D\ge 3$,  for the action to be physical, i.e. for its bosonic part to be quadratic in space time derivatives. In $D=2$ dimensions with $(p,q)$ supersymmetry, $D_-D_-\sim\partial _=$ and $D_+D_+\sim\partial _\+$ and a similar argument shows that  $p \le 2$ and $q \le 2$ for the action to be physical.

This bound on $d$ or $(p,q)$   
can be circumvented by   finding subspaces that are invariant under supersymmetry and integrating constrained Lagrangians over those.
The prime example of such subspaces are the chiral and antichiral subspaces of $D=4, ~{\cal N}=1$ superspace, where
the complex superfields $\phi$ obey the chirality condition $\bar D\phi=0$, 
and a chiral Lagrangian is 
 integrated with the chiral measure   $D^2$, and 
 an anti-chiral Lagrangian is  integrated with the anti-chiral measure 
 $\bar D^2$.
The projective superspace construction described in 
section \ref{proj}  below provides a systematic method of constructing such 
constrained superfields and the corresponding
invariant subspaces, but we first describe the approach of \cite{Gates:1984nk} .

\subsection{The GHR approach}
\label{GHR}

In \cite{Gates:1984nk}  a general invariant action for an off-shell $(4,4)$ multiplet was found. Here we adapt this to our $(4,1)$ models.

In constructing an action for $(4,1)$ multiplets we face the problem discussed above in section \ref{gen}.
The algebra involves four real or two complex positive chirality derivatives
$\bbD{+a}, \bbDB{+}^a$, and  so the full  $(4,1)$ superspace measure has too large a dimension. We then seek an invariant subspace and corresponding subintegration, similar to the chiral subspaces in ${\cal N}=1, D=4$ superspace. We use the procedure of \cite{Gates:1984nk}  and define two linear combinations of   positive chirality spinor derivatives:
\ber\nn\label{NAdefs}
&&\na_+=\beta\bbD{+1}+i\alpha\bbD{+2}\\[2mm]
&&\Delta_+={\alpha\bbDB{+}^1+i\beta\bbDB{+}^2}
\eer
for some choice of complex  parameters $\alpha , \beta$.

For a given choice of parameters $\alpha , \beta$, the $(4,1)$ superfields  
$\eta, \breve\eta$
given by
\ber \label{mult}
\label{etaetab}
 \eta:=\alpha \phi +\beta\bar\chi~,~~~\breve{\eta}:=\beta\bar\phi-\alpha \chi
\eer
are annihilated by $\na_+$ and $\Delta_+$
\ber\nn\label{etaetaba}
\na_+\eta =\Delta_+\eta=0~,~~~
\na_+\breve{\eta} =\Delta_+\breve{\eta}=0~.
\eer
Then for a Lagrangian constructed from these constrained superfields, a $(4,1)$  supersymmetric action is given using the conjugate  operators $\bar \na_+$ and $\bar\Delta_+$ to define the supermeasure.
The action is then
\ber\label{4,1}
i\int d^2x \bar \na_+\bar \Delta_+D_-L_-+h.c.~.
\eer
where  $h.c.$ denotes hermitian conjugate, and we take
\ber
L_-:=\lambda_i(\eta,\breve{\eta})D_-\eta^i+\tilde\lambda_i(\eta,\breve{\eta})D_-\breve{\eta}^i~,
\eer
for a set of multiplets labelled by the  index  $i$, for some potentials $\lambda_i,\tilde\lambda_i$.

A general action will be a linear superposition of actions of the form \re{4,1}. We then 
allow the potentials $\lambda_i,\tilde\lambda_i$ to depend explicitly on $\alpha,\beta$ and 
integrate  over all possible values of $\alpha,\beta$. 
The $(4,1)$  supersymmetric  action constructed from the constrained superfields in \re{etaetab} is then
\ber\nn\label{fullact}
&&i\int d^2x\left[\int d\alpha d\beta \bar \na_+\bar \Delta_+D_- L_-\right]~+h.c.\\[2 mm]
&&L_-:=\lambda_i(\eta,\breve{\eta}; \alpha,\beta)D_-\eta+\tilde\lambda_i(\eta,\breve{\eta}; \alpha,\beta)D_-\breve{\eta}~,
\eer
where the operators $\bar \na_+$ and $\bar\Delta_+$  define the supermeasure.
The parameter integration must be specified as some contour integral.

In the special case when the action is a  reduction of the $(4,4)$ action of   \cite{Gates:1984nk} which has a scalar function $L$ as its Lagrangian, one finds
\ber
-\tilde \lambda_i=\lambda_i=i\pa_{\eta^i+\breve{\eta}^i}L(\eta+\breve{\eta})~.
\eer

The measure in \re{4,1} can be rewritten in a form suitable for reduction to $(2,1)$ superspace using
\ber\label{NaDe}
\bar\Delta_+=-\frac{\bar \beta} \alpha \na_++\frac 1 \alpha (|\alpha |^2+|\beta |^2) \bbD{+1}~.
\eer
Since $\na_+$ and $\Delta_+$ annihilate the Lagrangian, the measure becomes
\ber\label{measure}
\bar\na_+\bar\Delta_+D_-\propto \bbD{+1}\bbDB{+}^1D_-~.
\eer
In the reduction we identify $\bbD{+1}\to\bbD{+}$ which gives the $(2,1)$ measure when the second $\theta^+$ is set to zero
\ber\label{measure2}
\bbD{+1}\bbDB{+}^1D_-(\dots)|=\bbD{+}\bbDB{+}D_-(\dots)|~.
\eer
This gives rise to an expression for the potential $k_\alpha$ in terms of an integral of an expression constructed from the $\lambda_i,\tilde \lambda_i $; we will give similar forms explicitly in later sections.
By construction, the potential $k_\alpha$ will necessarily satisfy the conditions \re{herm1},\re{herm2} for $(4,1) $ supersymmetry.

The form of $\eta,\breve{\eta}$ given in
\re{etaetab} implies that any function
 function $f(\eta,\breve{\eta})$ will automatically satisfy 
\ber\label{ohmy}
\frac{\pa^2f}{\pa\phi^i\pa\bar\phi^{\bar k}}+\frac{\pa^2f}{\pa\chi^k\pa\bar\chi^{\bar i}}=0~.
\eer
For the multiplet \re{mult}, this implies that the potential $k_\alpha$ constructed in this way
will satisfy
\ber
\label{asda}
k_{\alpha,\phi^k \bar \phi ^j}+k_{\alpha,\chi^j\bar \chi ^k}=0~,
\eer
and its complex conjugate, in addition to the conditions  \re{herm1},\re{herm2}  for $(4,1)$ supersymmetry. 

 Further, the potentials may be written
\ber\nn
&& k_{\phi^i}= i\left(\int d\alpha d\beta\alpha\lambda_i-\int d\bar\alpha d\bar\beta\bar\beta\bar{\tilde\lambda}_i\right) \\[1mm]
&& k_{\chi^i}= -i\left(\int d\alpha d\beta\alpha\tilde\lambda_i+\int d\bar\alpha d\bar\beta\bar\beta\bar{\lambda}_i\right)~,
\eer
along with their complex conjugates. Using this form, it is easy to show that the potentials actually satisfy the stronger conditions \re{vpotcs}. Thus the models constructed in this way 
 constitute a subclass of the possible $(4,1)$ models.

\section{Projective superspace}
\label{proj}
{The procedure  from \cite{Gates:1984nk} used in the derivation of the action \re{fullact} was introduced to construct an action for a particular multiplet. 
It was later realised that there is a generalisation that works the other way: 
the superspace can be enlarged by an extra coordinate or coordinates in such a way that 
superfields and actions in this enlarged superspace automatically 
  have extended supersymmetry. This is the  Projective Superspace construction
\cite{Karlhede:1984vr}, \cite{Lindstrom:1987ks}, \cite{Buscher:1987uw}, \cite{Lindstrom:1994mw}, a useful tool for finding new multiplets and constructing actions in various dimensions. We begin by making contact with the discussion in the previous section.}

\subsection{Relation of the GHR construction to Projective superspace.}

In the previous section we summed over theories parameterised by
complex  variables $(\alpha,\beta)$. The overall scale is unimportant, so they can be viewed as homogeneous coordinates on $  \mathbb{CP}^1$.
It is useful to instead use an inhomogeneous coordinate
 \ber 
 \zeta =i \alpha/\beta
 \eer
 in the region where $\beta \ne 0$, or 
 {$\zeta ' =-i \beta/ \alpha $} in the patch where $\alpha\ne 0$.
 Then the summation over theories corresponds to a contour integral on
 $  \mathbb{CP}^1$,{ covered by two patches, one with inhomogeneous coordinate $\zeta $ and 
 one with inhomogeneous coordinate $\zeta '$.
We now discuss Projective Superpace in more detail.}

\subsection{$(4,q)$ Projective superspace defined}

Projective  superspace is defined to deal with the limitations outlined in section \ref{gen} and at the same time gives a constructive method for finding new multiplets.
We shall be concerned with $(4,q)$ superspace for $q=4,2,1$. 
In all these cases a full superspace measure has more spinorial derivatives than allowed  {and so we seek invariant subintegrations.
Part of the construction is the same for all $p$, the difference is mainly in the form of the actions.

We start from the positive chirality part of the $D$ algebra given in the first line of \re{talg1} or \re{talg}.
A projective coordinate $\zeta$ on $  \mathbb{CP}^1$ is used to construct the combinations\footnote{The conventions have varied over time. The present choice are those of \cite{GonzalezRey:1997qh}, up to an unimportant overall $\zeta$ factor multiplying $\breve{\na}$ .}   
\ber\nn\label{defN}
&&\na_+:=\bbD{+1}+\zeta\bbD{+2}~,\\[1mm]
&&\breve{\na}_+:=\bbDB{+}^1-\zeta^{-1}\bbDB{+}^2~.
\eer
We introduce a conjugation acting on meromorphic functions of $f(\zeta)$  by

\ber
f(\zeta)\to \breve{f}(\zeta)
\eer
given by the composition of complex conjugation
\ber
f(\zeta)\to :f^*(\bar \zeta)\equiv ( f(\zeta))^*
\eer
and  the antipodal map
\ber
\zeta \to -\bar \zeta^{-1}
\eer
so that\footnote{Projective superspace uses complex conjugation composed with the antipodal map on $\mathbb{CP}^1$ \cite{Lindstrom:1987ks}, as described here. It is the relevant conjugation in projective superspace, and in the literature it is often denoted by just a bar. A closely related conjugation in harmonic superspace was earlier introduced in \cite{Galperin:1984av}.}
\ber
\breve{f}(\zeta)
=
 f^*(-\zeta^{-1}) ~.
 \eer
The derivatives  \re{defN} are related by the
this conjugation.}
We shall be interested in projectively chiral superfields $\eta$ that satisfy 
\ber\label{projchir}
\na_+\eta=0~,~~~\breve{\na}_+\eta=0~,
\eer
as well as being $(4,q)$ superfields.
We assume that they have the $\zeta$-expansion 
\ber
\eta=\sum_{\mu= -m}^n \zeta^\mu\eta_\mu~,
\eer
where $\eta_\mu$ is the expansion coefficient superfields for the $\mu$'th power of $\zeta$. {The constraints \re{projchir} then lead to the following conditions on the fields $\eta_\mu$:}
\ber\nn\label{cons}
&&\bbD{+1}\eta_\mu+\bbD{+2}\eta_{\mu-1}=0\\[1mm]
&&\bbDB{+}^1\eta_\mu-\bbDB{+}^2\eta_{\mu+1}=0~.
\eer
{{Here
$\eta_\mu=0$ for $\mu <-m$ and $\mu>n$, so that the highest and lowest components are constrained
\ber\nn\label{consasd}
&&\bbD{+1}\eta_ {-m} =0\\[1mm]
&&\bbDB{+}^1\eta_n=0~.
\eer}

To be able to write actions, two independent orthogonal derivatives are needed.
The following pair can be used for the supermeasure for fields annihilated by the operators (\ref{defN}):
\ber\nn\label{defM}
&&\Delta_+:=\bbD{+1}-\zeta\bbD{+2}~,\\[1mm]
&&\breve{\Delta}_+:=\bbDB{+}^1+\zeta^{-1}\bbDB{+}^2~.
\eer.

The algebra obeyed by the $\na$'s and $\Delta$'s is
\ber\nn\label{nadealg}
&&\{\na_+,\na_+\}=\{\breve{\na_+},\breve{\na_+}\}=\{\Delta_+,\Delta_+\}=\{\breve{\Delta}_+,\breve{\Delta}_+\}=\{\na_+,\Delta_+\}=\{\breve{\na}_+,\breve{\Delta}_+\}=0\\[1mm]
&&\{\na_+,\breve{\Delta}_+\}=\{\breve{\na}_+,{\Delta}_+\}=4i\pa_\+~.
\eer

\subsection{$(4,1)$ projective superspace}
For the  $(4,1)$ theories  the algebra is \re{talg1}.
The $(2,1)$ content of \re{cons} is then obtained as discussed previously  in Sec.\ref{Off}, by identifying the $(2,1)$ derivative as $\bbD{+}=\bbD{+1}$  and the generator of the non-manifest extra supersymmetries\footnote{See the comments following \re{8}. } as  ${\mathbb{Q}}_{+}=\bbD{+2}$. Most of the relations in \re{cons} will  just give the $\mathbb{Q}_+$ action of the second supersymmetry on the $\zeta$ coefficients fields $\eta_\mu$ . {Only the first and last fields in  the $\zeta$-expansion  in  \re{cons} will be constrained \footnote{  {We suppress the tildes that we previously used to denote  $(2,1)$ superfields.}}
\ber\nn
&&\bbD{+}\eta_{-m}=0\\[1mm]
&&\bbDB{+}\eta_n=0~.
\eer}
{The rest of the fields $\eta_\mu$  are unconstrained, with  the conditions \re{cons}
giving relations between $\eta_\mu$ and $\eta_{\mu\pm1}$.}
. 
A $(4,1)$ Lagrangian is   
\ber\label{exex}
i\oint_C\frac{d\zeta}{2\pi i \zeta} \Delta_+\breve{\Delta}_+D_-\left(\lambda_\alpha(\eta,\breve{\eta};\zeta)D_-\eta^\alpha+\breve{\lambda}_{\alpha}(\eta, \breve{\eta};\zeta)D_-\breve{\eta}^{\alpha}\right)~.
\eer
{The potentials $\lambda, \breve{\lambda}$ can depend explicitly on $\zeta$, and we perform a contour integration over a suitable contour $C$. In many examples, $C$ will be a small contour encircling the origin.}
Since it follows from \re{defN} and \re{defM} that $\Delta$ anticommutes with $D_-$, and that
\ber
\Delta_+=2\bbD{+}-\na_+~,
\eer
and since further $\na$ annihilates $\eta$ , we may make the following replacement in  reducing a Lagrangian  to $(2,1)$ superspace:
\ber\label{rep}
i\oint_C\frac{d\zeta}{2\pi i \zeta} \Delta_+\breve{\Delta}_+D_-{\cal L}_-(\eta,\breve{\eta})\to i\oint_C\frac{d\zeta}{2\pi i \zeta} \bbD{+}\bbDB{+}D_-{\cal L}_-(\eta,\breve{\eta})~.
\eer
 The relation of $\lambda_\alpha$ to $k_\alpha$ in \re{2,1} depends on the form of $\eta$, as illustrated in the examples below.
 
{After the reduction,  \re{rep} gives a $(4,1)$ supersymmetric   action written in $(2,1)$ superspace with the non-manifest supersymmetry ensured by the construction. 
 For the multiplet \re{etaetab},
 this will lead to constraints on ${\cal L}_- $ of the type \re{ohmy} . As before, these 
 lead to a potential $k$ satisfying \re{asda} in addition to the conditions  \re{herm1},\re{herm2}  for $(4,1)$ supersymmetry.
Thus for this multiplet,
 the models constructed in projective superspace  represent a subclass of the general $(4,1)$ models.
 
\subsubsection{Examples}
If we consider $\eta$'s with $m=0, n=1$,  and denote  $\eta_0=\bar \phi$, $\eta_1=\chi$, we have
\ber\label{simpeta}\nn
&&\eta^i=\bar\phi^i+\zeta \chi^i\\[1mm]
&&\breve{\eta}^i=\phi^i-\zeta^{-1}\bar\chi^i~,
\eer
{with $i=1\dots d$ for $d$ fields $\eta^i$.}
From \re{cons} we find that the coefficients obey
\ber\nn
&&\bbDB{+} \phi^i=0~,~~~\bbDB{+}\chi^i=0~,~~~{\mathbb{Q}}_+\phi^i=0~,~~~{\mathbb{Q}}_+\chi^i=0\\[1mm]
&&\bar{\mathbb{Q}}_+\phi^i=-\bbDB{+}\bar\chi^i~,~~~\bar{\mathbb{Q}}_+\chi^i=\bbDB{+}\bar\phi^i~.
\eer 
For each $i$, this is \re{constr2} with $i\bbD{+2}=\mathbb{Q}_+$. From \re{exex}, the $(2,1)$ Lagrangian is 
\ber\label{etact1}
i\oint_C\frac{d\zeta}{2\pi i \zeta} \bbD{+}\bbDB{+}D_-
\left(\lambda_{\eta^i}
(\eta,\breve{\eta})D_-\eta^i
    +\breve{ \lambda}_{\breve{\eta}^i}
(\eta,\breve{\eta})
D_-{\breve{\eta}}^i \right)~,
\eer
In this case, the relation of $\lambda_i$ to $k_i$ in \re{2,1} is given by\footnote{Note that the $\zeta$ measure is invariant under conjugation.}
\ber\nn\label{ks}
&&k_{\phi^i}=\oint_C\frac{d\zeta}{2\pi i \zeta}\breve{\lambda}_i~,~~~\bar k_{\bar\phi^i}=\oint_C\frac{d\zeta}{2\pi i \zeta}{\lambda}_i\\[1mm]
&&k_{\chi^i}=\oint_C\frac{d\zeta}{2\pi i \zeta}\zeta{\lambda}_i~,~~~~\bar k_{\bar\chi^i}=-\oint_C\frac{d\zeta}{2\pi i \zeta}\zeta^{-1}\breve{\lambda}_i~.
\eer
By construction, these potentials satisfy \re{asda}
as well as
 \re{herm1} and \re{herm2}. In fact, again, a direct calculation using \re{ks} shows  they satisfy the stronger condition \re{vpotcs}. As a result,  the Lagrangian \re{exex} is not the most general one with $(4,1)$ supersymmetry.
 
 To see that the vector potentials in \re{ks} satisfy  \re{vpotcs}, we form their derivatives, using \re{simpeta},
 \ber\nn
 &&k_{\phi^i,\bar\phi^j}=\oint_C\frac{d\zeta}{2\pi i \zeta}\breve{\lambda}_{i,\eta^j}~,~~~\bar k_{\bar\chi^i,\chi^j}=-\oint_C\frac{d\zeta}{2\pi i \zeta}\zeta^{-1}\breve{\lambda}_{i,\eta^j}\zeta.\\[1mm]
 &&k_{\phi^i,\bar\chi^j}=-\oint_C\frac{d\zeta}{2\pi i \zeta}\breve{\lambda}_{i,\eta^j}\zeta^{-1}~,~~~\bar k_{\bar\chi^i,\phi^j}=-\oint_C\frac{d\zeta}{2\pi i \zeta}\zeta^{-1}\breve{\lambda}_{i,\eta^j}~.
 \eer
They clearly satisfy  \re{vpotcs}.

Consider now the example of a quadratic Lagrangian  for $d$  multiplets ${\eta}^i$ given by
\ber
{\cal L}_-=i\oint_C\frac{d\zeta}{2\pi i \zeta}\left(\eta^iD_-\breve{\eta}^i-\breve{\eta}^iD_-\eta^i\right)~,
\eer 
where the contour $C$ is a small circle around the origin. Using (\ref{simpeta})
and
 performing the $\zeta$ integration results in the following  action
\ber\label{ex103}
&&-\int d^2x\bbD{+}\bbDB{+}D_-\left(\bar\phi^i D_-\phi^i+\bar\chi^iD_-\chi^i-\phi^i D_-\bar\phi^i-\chi^iD_-\bar\chi^i\right)~
\eer
with
\ber
k_{\phi^i}=i\bar\phi^i~, ~~~k_{\chi^i}=i\bar\chi^i~,
\eer
in agreement with \re{ks}.

A more interesting example arises if we take a general real function $L(\eta+\breve{\eta})$ and set 
$\lambda=-\breve{\lambda}=iL$. . The vector potentials may be  immediately read off from \re{ks} using these expressions:
\ber\nn\label{ks1}
&&k_{\phi}=\oint_C\frac{d\zeta}{2\pi i \zeta}\breve{\lambda}=-iL_0~,~~~~~\bar k_{\bar\phi}=\oint_C\frac{d\zeta}{2\pi i \zeta}{\lambda}=iL_0\\[1mm]
&&k_{\chi}=\oint_C\frac{d\zeta}{2\pi i \zeta}\zeta{\lambda}=iL_{-1}~,~~~~\bar k_{\bar\chi}=-\oint_C\frac{d\zeta}{2\pi i \zeta}\zeta^{-1}\breve{\lambda}=iL_1~,
\eer
where $L_\mu$ are the coefficients in a expansion of $L$ in powers of $\zeta$.
This will lead to a metric $g$ and $B$ field given by the $\zeta^1, \zeta^0$ and $\zeta^{-1}$ components of the derivative of $L$ according to
\ber
E=g+B=\left(\begin{array}{cccc}
0&L'_0& L'_{-1}&0\\
L'_0&0&0&-L'_1\\-L'_{-1}&0&0&L_0'\\
0&L'_1&L'_0&0
\end{array}\right)~,
\eer
with prime denoting derivative with respect to the argument and rows and columns ordered as $(\phi,\bar\phi,\chi,\bar\chi)$. As an example, a function 
\ber
L=\frac 1 {3!} (\eta+\breve{\eta})^3
\eer
gives
\ber\nn
&&g_{\phi\bar\phi}=g_{\chi\bar\chi}=(\phi+\bar\phi)^2-2\chi\bar\chi\\[1mm]\nn
&&B_{\phi\chi}=-2(\phi+\bar\phi)\bar\chi~,\\[1mm]
&&B_{\bar\phi\bar\chi}=-2(\phi+\bar\phi)\chi~.
\eer

An example involving unconstrained fields arises when $m\ne 0$. We then consider
\ber\label{genex}
\eta=\sum_{-m}^n\zeta^\mu\eta_\mu~
\eer
where the top coefficient $\eta _n\equiv \chi$ and the bottom component  $\eta_{-m}\equiv\bar\phi$ give   chiral  fields $\phi, \chi$ in the $(2,1)$ reduction, while the rest of the fields $\eta^i_\mu$
for $ -m<\mu <n$
 are unconstrained. The $(4,1)$ transformations that follow from the constraints are 
\ber\nn
&&\bbD{+} \eta_{-m}=\bbD{+} \bar \phi=0~, ~~~\bbDB{+}  \eta_{n}=\bbDB{+} \chi=0\\[1mm]\nn
&&\bar{\mathbb{Q}}_+\eta_{\mu+1}=\bbDB{+} \eta_{\mu}~,~~~{\mathbb{Q}}_+\eta_{\mu-1}=-\bbD{+} \eta_{\mu}~,~~~\mu=n-1,...,-m+1\\[1mm]
&&\bar{\mathbb{Q}}_+\eta_{-m}=\bar{\mathbb{Q}}_+\bar\phi=0~,~~~{\mathbb{Q}}_+ \eta_n={\mathbb{Q}}_+ \chi=0~.
\eer

This last example goes beyond the models described by the action \re{2,1}, and introduces new unconstrained superfields.
In particular, consider the following $\eta$ with $m=1=n$:
\ber\nn
&&\eta=\zeta^{-1}\bar\phi+X+\zeta \chi\\[1mm]
&&\breve{\eta}=-\zeta\phi+\bar X -\zeta^{-1}\bar\chi~,
\eer
The fields satisfy
\ber\nn
&&\bbD{+} \bar \phi=0~, ~~~\bbDB{+} \chi=0~,\\[1mm]\nn
&&\bar{\mathbb{Q}}_+ X=\bbDB{+} \bar \phi~,~~~{\mathbb{Q}}_+X=-\bbD{+} \chi~,\\[1mm]\nn
&&{\mathbb{Q}}_+\bar\phi=-\bbD{+} X~, \bar {\mathbb{Q}}_+\chi=\bbDB{+} X\\[1mm]
&&\bar{\mathbb{Q}}_+\bar\phi=0~,~~~{\mathbb{Q}}_+ \chi=0~,
\eer
which leaves $X$ unconstrained.  A quadratic Lagrangian is 
\ber
{\cal L}_-=i\oint_C\frac{d\zeta}{2\pi i \zeta}\left(\eta D_-\breve{\eta}-\breve{\eta}D_-\eta\right)~.
\eer
Performing the $\zeta$ integration results in the $(2,1)$ action taking the form
\ber\label{flatact}
S=-\int d^2x\bbD{+}\bbDB{+}D_-\left(\bar\phi D_-\phi-\phi D_-\bar\phi+XD_-\bar X-\bar XD_-X-\bar\chi D_-\chi+\chi D_-\bar\chi\right)~.
\eer
The superfields $\phi,\chi$ are chiral and satisfy the standard free field equations
\be
\bbD{+} D_- \phi=0, 
\qquad
\bbD{+} D_- \chi=0~.
\ee
However, note that their kinetic terms in the action have opposite sign.
The superfield
$X$ is unconstrained and its
  field equation is $D_-X=0$, which implies 
$  \partial _=X=0$.
The components $X \vert, \bbD{+}X \vert, \bbDB{+}X \vert, \bbD{+}\bbDB{+}X \vert$
are all right-moving, i.e. are independent of $x^=$, while the remaining components
 $D_-X \vert, \bbD{+}D_-X \vert, \bbDB{+}D_-X \vert, \bbD{+}\bbDB{+}D_-X \vert$
 are all set to zero by the field equations.

\subsection{$(4,2)$ projective superspace}
\label{4.2}
For   $(4,2)$ {superspace} 
the derivative algebra is \re{talg}:
\ber\nn
&&\{\bbD{+a},\bbDB{+} ^b\}=~2i\delta^b_a\pa_\+~, ~~~a,b,=1,2.\\[1mm]
&&\{\bbD{-},\bbDB{-}\}=2i \pa_=~.
\eer
As before, we introduce  projectively chiral superfields $\eta$, now in $(4,2)$ superspace,  that satisfy 
\ber\label{projchir1}
\na_+\eta=0~,~~~\breve{\na}_+\eta=0~,
\eer
which have the $\zeta$-expansion 
\ber
\eta=\sum_{-m}^n \zeta^\mu\eta_\mu~.
\eer
In addition, we impose  chirality constraints, to obtain irreducible multiplets
\ber\label{dminus}
\bbD{-}\eta=0~,~~\Rightarrow ~~~\bbDB{-}\breve{\eta}=0~.
\eer
Then
the top coefficient $\eta _n\equiv \chi$ and the bottom component  $\eta_{-m}\equiv\bar\phi$ give      fields $\phi, \chi$ in the $(2,2)$ reduction, where $\phi$ is chiral; and $\chi$ is twisted chiral.

 An invariant action is
\ber\nn \label{actar}
&&\oint_C\frac{d\zeta}{2\pi i \zeta}\Delta_+\bar\Delta_+\bbD{-}\bbDB{-}L(\eta,\breve{\eta}:\zeta)=\oint_C\frac{d\zeta}{2\pi i \zeta}\bbD{+}\bbDB{+}\bbD{-}\bbDB{-}L(\eta,\breve{\eta}:\zeta)\\[2mm]
&&=:\bbD{+}\bbDB{+}\bbD{-}\bbDB{-}K|
\eer
where $L$ and its $\zeta$ integral $K$  are  real potentials. Note that terms of the form $f(\eta)+\breve{f}(\breve{\eta})$ integrate to zero in the action and thus 
{shifts $L \to L+ f(\eta)+\breve{f}(\breve{\eta})$}
constitute ``K\"ahler gauge transformations''.
The non-manifest supersymmetry transformations are 
\ber
\bar{\mathbb{Q}}_+\eta=\zeta\bbDB{+}\eta~.
\eer

The reduction to $(2,2)$ superspace \re{actar} gives a potential $K$ which
automatically satisfies 
\ber\label{ohmysd}
\frac{\pa^2K}{\pa\phi^i\pa\bar\phi^{\bar k}}+\frac{\pa^2K}{\pa\chi^k\pa\bar\chi^{\bar i}}=0~.
\eer
This is precisely the condition \re{4.1cnd} for $(4,2)$ supersymmetry, so in this case projective superspace gives the most general $(4,2)$ supersymmetric model.

Note that a variant multiplet $\tilde \eta$ arises if we replace \re{dminus} by
\ber\label{dminus2}
\bbDB{-}\hat\eta=0~,~~\Rightarrow ~~~\bbD{-}\breve{\hat\eta}=0~
\eer
which corresponds to $\theta_1^- \leftrightarrow \bar\theta^{1-}$. However, it is easy to see that $\hat \eta(\bar\phi,\chi) $ is equivalent to $ \breve{\eta}(-\bar\phi,\chi)$  for $\eta=\bar\phi+\zeta\chi$.

\subsubsection{Example}

A simple example of  a $(4,2)$ multiplet is 
\ber\label{42mult}
&&\eta=\bar\phi+\zeta\chi~.
\eer
The
projective chirality constraints {result in $(2,2)$ superfields $\phi,\chi$ 
with $\phi$ chiral and $\chi$  twisted chiral.} They also yield the transformations 
\ber\label{2var}
&&\bar{\mathbb{Q}}_+\bar\phi=0~,~~~\bar{\mathbb{Q}}_+\bar\chi=0~,~~~\bar{\mathbb{Q}}_+\chi=\bbDB{+}\bar\phi~,~~~~\bar{\mathbb{Q}}_+\phi=-\bbDB{+}\bar\chi
\eer
corresponding to the truncation of the $(4,4)$ multiplet. The formulae in \re{2var} are those in \re{constr21} 
$\mathbb{Q}_+ \to i\bbD{+2}$.
We obtain the scalar potential
\ber
K=i\oint_C\frac{d\zeta}{2\pi i \zeta}L~.
\eer
From the action \re{actar}, reducing to $(2,1)$ and setting $\lambda_{\eta^i} \to -iL_{,\eta^i}$ we find
\ber\nn
&&k_{\phi^i}=i\oint_C\frac{d\zeta}{2\pi i \zeta}L_{,\breve{\eta}^i}~,~~~\bar k_{\bar\phi^i}=-i\oint_C\frac{d\zeta}{2\pi i \zeta}L_{,{\eta}^i}\\[1mm]
&&k_{\chi^i}=-i\oint_C\frac{d\zeta}{2\pi i \zeta}\zeta L_{,{\eta}^i}~,~~~\bar k_{\bar\chi^i}=-i\oint_C\frac{d\zeta}{2\pi i \zeta}\zeta^{-1}L_{,\breve{\eta}^i}~.
\eer

\subsubsection{Flat space}
In particular, considering a quadratic function of  $d$  multiplets
\ber
L=\eta^i\breve{\eta}^i~,
\eer with the contour $C$ a small circle around the origin, gives the following $(2,1)$ action
\ber\nn
&&\int d^2x\oint_C\frac{d\zeta}{2\pi i \zeta}\bbD{+}\bbDB{+}D_-\left(\breve{\eta}^iD_-\eta^i-\eta^iD_-\breve{\eta}^i\right)~.
\eer
Performing the $\zeta$ integration reproduces the component action \re{ex103}, but with the fields now being chiral and twisted chiral. The target space geometry is $2d$ dimensional, flat with zero $B$-field.  

\subsubsection{Curved space}
We can construct a more interesting model using  the propeller contour $\Gamma$ in Fig.1; a similar construction was used in \cite{Karlhede:1984vr}.  
\captionsetup{width=0.8\textwidth}
\begin{figure}[!ht]
  \centering
    \includegraphics[width=0.6\textwidth=1.0]{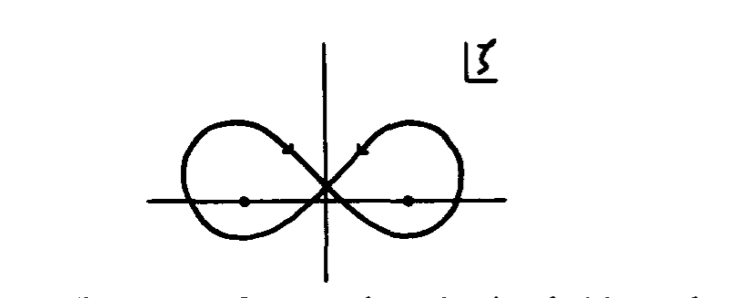}
    \caption{A propeller contour encircling two singularities of $ln~\!(\bullet )$. The zeros are depicted as lying on the real axis, but in our example they lie on a line tilted to an angle $\theta$ with the real axis, see \re{zeros}.} 
\end{figure}
We use the
the $(4,2)$ multiplet $\eta$ in \re{42mult} and consider the Lagrangian given by the following integral over the contour $\Gamma$:
\ber
-\oint_\Gamma \frac{d\zeta}{2\pi i \zeta}(\eta+\breve{\eta})~\!\! ln~\!\! (\eta\breve{\eta} )~.
\eer
Regarding $\eta\breve{\eta}$ as a function of $\zeta$, it has two zeroes
\ber\label{zeros}
\zeta_1= -\frac{\bar \phi} \chi=:-\frac 1 r e^{i\theta}~,~~~~\zeta_2= \frac{\bar \chi}\phi =: r e^{i\theta}~,
\eer
and these are branch points of $ln (\eta\breve{\eta})$. We take one branch cut to go from $\zeta _1$ to $-\infty$ on the real axis, and the other to go from $\zeta _2$ to $+\infty$ on the real axis.
For any $f(\zeta) $, the integral
$$\frac 1 {2\pi i  } \, \int _\Gamma{d\zeta} f(\zeta) ln (\eta\breve{\eta})$$
gives 
 the definite integral 
 $$\int _{\zeta_1}^{\zeta_2}{d\zeta} f(\zeta) $$
along the straight line between  $\zeta_1$ and $\zeta_2$.
\footnote{ {This can be seen as follows. The real and imaginary axes divide the $\zeta$-plane into four quadrants.
 Choose a branch where the integral is $\int {d\zeta} f(\zeta) ln (\eta\breve{\eta})$ along the part of the curve below the negative real axis,
 i.e. in the bottom left quadrant.
  Above the negative and below the positive real axes (i.e. in the upper left and lower right quadrants)
  we then have $\int {d\zeta} f(\zeta) (ln(\eta\breve{\eta}) +2\pi i)$ and above the positive real axis (i.e. in the upper right quadrant), changing sheet in the opposite direction, it is   $\int {d\zeta} f(\zeta) ln (\eta\breve{\eta})$ again. Combining the integrals and paying attention to  the directions of integration the net result is $\int _{\zeta_1}^{\zeta_2} d\zeta f(\zeta)$. 
  }}

For $f(\zeta)= \frac{1}{ \zeta}(\eta+\breve{\eta})$,
the resulting $(2,2)$ scalar potential is then
\ber
K=-(\phi+\bar\phi)\left( \frac{\chi\bar\chi}{\phi\bar\phi}+ln~\!(\frac{\chi\bar\chi}{\phi\bar\phi})\right)~,
\eer
(up to K\"ahler gauge transformations), which is indeed invariant under the additional supersymmetry in \re{2var}. The geometry has a conformally flat metric $g$  and  is given by 
\ber\nn
&&g_{a\bar b}=\left(\begin{array}{cc}g_{\phi\bar\phi}&0\\
0&g_{\chi\bar\chi}\end{array}\right)=\frac{(\phi+\bar\phi)}{\phi\bar\phi}\left(\begin{array}{cc}\mathbb{1}&0\\
0&\mathbb{1}\end{array}\right)\\[2mm]\nn
&&H=\phi^{-2}d\chi\wedge d\bar\chi\wedge d\phi-\bar\phi^{-2}d\chi\wedge d\bar\chi\wedge d\bar\phi\\[1mm]
&& R= {\frac 3 2}
\frac {\phi\bar\phi} {(\phi+\bar\phi)^3} 
\eer
where $H=dB$ and $R$ is the curvature scalar (see, e.g.,  \cite{Brendle}).
Note  the vector field $\partial/\partial \chi$ generates an isometry. 


\subsubsection{Semichiral Superfields}

When we want to generalise the constructions along the lines of the second $(4,1)$ example \re{genex} above, we run into an interesting problem. When $n\leq 1$, and $\eta$ is a series as in \re{genex} with the additional condition $D_-\eta=0$, { its  reduction to $(2,2)$ superspace contains right semichiral fields rather than unconstrained superfields, } e.g.,
\ber
\eta=\zeta^{-1}\eta_{-1}+\eta_0+\zeta\eta_1 = \zeta^{-1}\bar\phi+\bar r +\zeta\chi~,
\eer
the constraints imply  that $\phi$ is chiral, $\chi$ twisted chiral and $r$ right semichiral: $\bbDB{-}r=0$. {However,   to construct a sigma model with a non-degenerate kinetic term, one needs an equal number of left and right semichiral superfields. Here by necessity we get right semichiral superfields but no left semichiral superfields. Such a model typically contains right-moving multiplets
\cite{Buscher:1987uw}. The  $(4,2)$ projective superfields are  thus restrictive when it comes to constructing sigma models.
To construct a sigma model with non-degenerate kinetic term, the multiplets considered here would need to be combined with other multiplets.}

\subsection{$(4,4)$ projective superspace}

This case is well documented in the literature \cite{Karlhede:1984vr}-\cite{Lindstrom:2008gs}, {and we make no claim of completeness for  the following brief presentation. The construction is always off-shell, typically involving auxiliary fields (sometimes an infinite number).} The application to our present type of multiplets, notably the $(4,4)$ twisted multiplet, requires use of the doubly projective superspace based on $\mathbb{CP}^1 \otimes\mathbb{CP}^1$. The two coordinates on these are labeled $\zeta_L$ and $\zeta_R$, respectively. The linear combinations of the four $(4,4)$ derivatives are
\ber\nn
&&\na_+:=\bbD{+1}+\zeta_L\bbD{+2}~, ~~~\na_-:=\bbD{-1}+\zeta_R\bbD{-2}\\[1mm]
&&\Delta_+=\bbD{+1}-\zeta_L\bbD{+2}~, ~~~\Delta_-=\bbD{-1}-\zeta_R\bbD{-2}~,
\eer 
and their 
conjugates. Now the anticommutation relations are \re{nadealg} for 
positive chirality derivatives $\na_+,\Delta_+$
with  similar relations for the negative chirality ones $\na_-,\Delta_-$.
A projectively chiral superfield  $\eta$ satisfies
\ber\label{fetadef}
\na_\pm \eta=0~.
\eer
We consider the real multiplet
\ber
\eta=\bar\phi+\zeta_L\chi+\zeta_R\bar\chi-\zeta_L\zeta_R\phi~,
\eer
where the components and transformations are those of the $(4,4)$ twisted multiplet and the reality condition is
\ber
\eta=-\zeta_L^{-1}\zeta_R^{-1}\breve{\eta}~.
\eer
A $(4,4)$ Lagrangian is
\ber
\oint _{C_L}\frac{d\zeta_L}{2\pi i \zeta_L}\oint_{C_R} \frac{d\zeta_R}{2\pi i \zeta_R} \Delta_+\Delta_-\breve{\Delta}_+\breve{\Delta}_-L(\eta)~,
\eer
where  $C_L$ and $C_R$ are some suitable contours. 
By construction, 
this will be invariant under the full $(4,4)$ supersymmetry. In fact, $L=L(\eta^i)$ ensures that the potential $K$ satisfies the general $(4,4)$ conditions \re{ertyb} and \re{ertyc}, where the indices now refer to a set of $\eta^i~\!\!$s.

Other multiplets involving semichirals and auxiliaries may be constructed as in \cite{Buscher:1987uw} and  \cite{Lindstrom:1994mw}.

{Finally, we mention that other extended superspaces, such as Harmonic Superspace \cite{Galperin:1984av}-\cite{Galperin:2001uw}, have also been used to describe off-shell $(4,4)$ multiplets and actions.  The construction closest to what we describe in this section uses {bi-harmonic superspace, as described in, e.g.,  \cite{Ivanov:1995jb}.}


\bigskip
{\section{Conclusion}

In this paper we introduce new $(4,1)$ and $(4,2)$ multiplets and construct actions for them  using new projective superspaces and their progenitors in the GHR formalism. We find the conditions for additional supersymmetries as conditions on the geometric objects: the vector or scalar potentials for the metric and $B$-field. Our multiplets and actions display off-shell supersymmetry and simultaneously integrable complex structures.

The general conditions for a $(2,1)$ model to have $(4,1)$ symmetry are given in  \re{herm1} and \re{herm2}. The conditons for a $(2,2)$ model to have $(4,2)$ symmetry are \re{4.1cnd}, and the conditons for a $(2,2)$ model to have $(4,4)$ symmetry are the well known relations \re{ertyc} and \re{ertyb}.
We also consider a stronger condition \re{vpotcs} that is sufficient but not necessary for a $(2,1)$ model to have $(4,1)$ symmetry.

Actions for the $(4,1)$ multiplet \re{constr2} as well as for $(4,2)$ multiplets are constructed both using the GHR approach and  novel $(4,1)$ and $(4,2)$ projective superspaces. 

We briefly reviewed the $(4,4)$ models.  General $(4,4)$ models were formulated in $(4,4)$ superspace  using the GHR approach in \cite{Gates:1984nk} later  using  projective superspace actions. In both approaches,   the scalar potential satisfies certain conditions by construction. These full conditions for $(4,4)$ supersymmetry arise when we combine the conditions for $(4,2)$ with the conditions for $(2,4)$, supersymmetry.

Examining the $(4,p)$ supersymmetric actions constructed in $(4,p)$ superspace using both the projective superspace and GHR constructions, we find that 
they give the most general $(4,p)$ supersymmetric sigma models for both the $(4,2)$ and $(4,4)$ cases, but for the $(4,1)$ case we obtain only the special class of models for which the constraint \re{constr2} is satisfied.
This can be viewed as follows.
The $(4,1)$ actions we have constructed are based on superfields  that depend on additional parameters apart from the worldsheet superspace  coordinates. The  additional  parameters enter in such a way that the second derivative conditions  \re{ohmy} are satisfied. In addition to this, the form of the actions leads to vector potentials that satisfy \re{vpotcs}. Together these conditions are stronger than the general conditions \re{herm1} and \re{herm2} for extra supersymmetry of a $(2,1)$ action. This is in contrast to the $(4,2)$ case where the conditions derived for the scalar potential that depends on extra parameters satisfies the general condition \re{4.1cnd} for $(4,2)$ supersymmetry. At present we do not fully understand this discrepancy, but perhaps there is a more general construction which gives a manifest formulation of the general $(4,1)$ case.

\bigskip

\noindent{\bf Acknowledgement}:\\
During the final stage of writing this article, we were informed of interesting results by  Naveen Prabhakar and Martin Ro\v cek on $(0,4)$ and $(0,2)$ multiplets and projective superspace that partly overlap, but mostly complement our presentation. We hope to return to the topic in a joint publication.
Discussions with both these authors, as well as the stimulating atmosphere at the 2016 Simons workshop on geometry and physics, are gratefully acknowledged. UL gratefully acknowledges the hospitality of the theory group at Imperial College, London, as
well as partial financial support by the Swedish Research Council through VR grant 621-2013-4245.This work was supported  by the EPSRC programme grant "New Geometric
Structures from String Theory", EP/K034456/1.

\appendix
\section{Connections}
\label{app1}

In the $(2,1)$ formulation one complex structure, $\mathbb{J}^{(1)}$  has its canonical form and is preserved by a connection with torsion. The form  of $\mathbb{J}^{(1)}$ follows from the reduction of the $(2,1)$ constraint
\ber
\bbDB{+}\varphi^\alpha=0
\eer
to $(1,1)$ as in \re{8}:
\ber\nn
\bbD{+}=: D_+-i\check{D}_+~,
\eer
which implies that
\ber
{\check{D}_+ \left(\begin{array}{c}\varphi^\alpha\\
\bar\varphi^{\bar \alpha}\end{array}\right) } \Bigg|
=Q_+ \left(\begin{array}{c}\varphi^\alpha\\
\bar\varphi^{\bar \alpha}\end{array}\right)=\mathbb{J}^{(1)}\left(\begin{array}{c}\varphi^\alpha\\
\bar\varphi^{\bar \alpha}\end{array}\right)=\left(\begin{array}{cc}i\mathbb{1}&0\\
0&-i\mathbb{1}\end{array}\right)D_+\left(\begin{array}{c}\varphi^\alpha\\
\bar\varphi^{\bar \alpha}\end{array}\right)~.~
\eer
 Invariance of the action implies
\ber\label{consJ}
\na^{(+)}\mathbb{J}^{(1)}=0~.
\eer

For the torsion-free case, $\na^{(0)}\mathbb{J}^{(1)}=0$ implies that the Levi-Civita connection has no mixed ``holonomy'' components, i.e. $\Gamma_{i~ \bar \beta}^{~\alpha}=0$.
SImilarly,  for the  connection $\Gamma^{(+)}$ with torsion
\ber T_{ij}{}^k= \frac 1 2 g^{kl} H_{ijl}
\eer
\re{consJ} 
implies the connection $\Gamma^{(+)}$ has no mixed ``holonomy'' components,
so that\ber\label{torrel}
\Gamma^{(+)~\!\!\alpha}_{i~\!\bar \alpha}=\Gamma^{(0)~\!\!\alpha}_{i~\!\bar \alpha}+T^{~~\alpha}_{i~\!\!\bar \alpha}=0~,~~~\Gamma^{(+)~\!\!\bar \alpha}_{i~\!\alpha}=\Gamma^{(0)~\!\!\bar \alpha}_{i~\!\alpha}+T^{~~\bar \alpha}_{i~\!\!\alpha}=0~.
\eer
In addition, the hermiticity condition 
\ber
\mathbb{J}^{(1)t}g\mathbb{J}^{(1)}=g~,
\eer
implies that the metric  has only mixed components, $g_{\alpha\beta }=0$,
and this determines the Levi Civita connections:
\ber\nn\label{lc}
&&\Gamma^{(0)}_{~\!\!\alpha~\!\!\beta~\!\!\gamma}=0~,~~~\Gamma^{(0)}_{~\!\!\bar \alpha~\!\!\bar \beta~\!\!\bar \gamma}=0\\[1mm]\nn
&&\Gamma^{(0)}_{~\!\!\alpha~\!\!\beta~\!\!\bar \gamma}=g_{\bar \gamma(\alpha,\beta)}~,~~~\Gamma^{(0)}_{~\!\!\bar \alpha~\!\!\bar \beta~\!\!\gamma}=g_{\gamma(\bar \alpha,\bar \beta)}\\[1mm]
&&\Gamma^{(0)}_{~\!\!\alpha~\!\!\bar \beta~\!\! \gamma}=\Gamma^{(0)}_{~\!\!\bar \beta~\!\!\alpha~\!\! \gamma}=g_{\bar \beta [\gamma,\alpha]}~,~~~\Gamma^{(0)}_{~\!\!\alpha~\!\!\bar \beta~\!\! \bar \gamma}=\Gamma^{(0)}_{~\!\!\bar \beta~\!\!\alpha~\!\! \bar \gamma}=-g_{\alpha [\bar \beta,\bar \gamma]}~.
\eer
where for any connection we define
\ber
\Gamma_{ijk} =g_{kl} \Gamma _{ij}^l
\eer

The formulae \re{gB} for the metric and B-field imply that 
\ber\nn\label{gbrel}
&&g_{\bar \beta [\gamma,\alpha]}=i(\bar k_{\bar \beta,[\gamma}-k_{[\gamma,\bar \beta})_{,a]}=-ik_{[\gamma,\alpha]\bar \beta}=\half B^{(2,0)}_{\gamma\alpha\bar \beta}=\half H^{(2,1)}_{\gamma\alpha\bar \beta}=T_{\gamma\alpha\bar \beta}~,\\[1mm]
&&-g_{\alpha [\bar \beta,\bar \gamma]}=\half B^{(0,2)}_{\bar \beta \bar \gamma,\alpha}=\half H^{(1,2)}_{\bar \beta \bar \gamma\alpha}=T_{\bar \beta \bar \gamma\alpha}~.
\eer
Combining \re{lc} and \re{gbrel} we see that the relations in \re{torrel} are satisfied.

The connection with torsion is then given by
\ber
\Gamma^{(+)}_{~\!\!\alpha~\!\!\beta~\!\!\bar \gamma}=
\Gamma^{(0)}_{~\!\!\alpha~\!\!\beta~\!\!\bar \gamma}+T_{~\!\!\alpha~\!\!\beta~\!\!\bar \gamma}=g_{\bar \gamma(\alpha,\beta)}+
g_{\bar \gamma[\alpha,\beta]}=
g_{\bar \gamma\alpha,\beta}
\eer
and
\ber
\Gamma^{(+)}_{~\!\!\alpha~\!\!\bar \beta~\!\! \gamma}=
\Gamma^{(0)}_{~\!\!\alpha~\!\!\bar \beta~\!\! \gamma}+T_{~\!\!\alpha~\!\!\bar \beta~\!\! \gamma}=2g_{\bar \beta [\gamma,\alpha]}
\eer
together with
\ber
\Gamma^{(+)}_{~\!\!\alpha~\!\! \beta~\!\! \gamma}=
\Gamma^{(+)}_{~\!\!\alpha~\!\!\bar \beta~\!\! \bar\gamma}=0
\eer


   \end{document}